\documentstyle[11pt]{article}

\begin{document}

\thispagestyle{empty}

\begin{center}
{\Large {\bf Higher-shell corrections for systems with one and
two valence nucleons: A spin-orbit and tensor interaction analysis}}

\vspace{0.3in}

D. C. Zheng

\begin{small}
{\it Department of Physics, University of Arizona, Tucson, AZ 85721}
\end{small}

\vspace{0.2in}

L. Zamick and M. Fayache

\begin{small}
{\it Department of Physics and Astronomy, Rutgers University,
	Piscataway, NJ 08855}
\end{small}

\end{center}

\begin{abstract}
It is shown that when higher-shell admixtures are included for
systems with two valence particles or holes,
there are effects which are quite different from those
for one-valence-nucleon systems. For example,
for nuclei with one valence particle or hole,
there is no first-order correction for the magnetic dipole moment or
the Gamow-Teller transition amplitude.
However for nuclei with two valence particles or holes,
one can get substantial corrections.
The effects of the tensor and spin-orbit interactions in core
renormalization are emphasized. We find that
in $^6\mbox{Li}$, the spin-orbit interaction causes the quadrupole
moment of the $J^{\pi}$=$1^+$ ground state to be positive, but the
tensor interaction causes it to be negative.
The G-matrices derived from realistic interactions are employed in these
calculations.
\end{abstract}

\pagebreak

\section{Introduction}
There has been considerable study of core renormalization
for a system of a closed $LS$ shell plus or minus one nucleon.
The most studied problem is probably the
electric quadrupole (E2) effective charge.
When doing calculations of E2 properties, i.e., quadrupole moments
or $B({\rm E2})$'s in a single major shell, it is necessary to use
effective charges for the valence
protons and neutrons. The quadrupole moment for the
ground state of $^{17}\mbox{O}$ is about --3$e\, {\rm fm}^2$,
suggesting that the effective charge for the valence $d_{5/2}$-neutron
is about -3/(-6)=0.5, where the quadrupole moment for a
$d_{5/2}$-proton is taken to be --6$e\, {\rm fm}^2$.
The same analysis would lead to an effective charge of 1.2
for a valence proton, although more careful considerations may lead
to somewhat different values. The use of effective charges can be
partly justified in first-order perturbation theory by
the core-polarization process shown in
Fig.1(a). Larger effective charges can be obtained in R.P.A.
calculations \cite{sz}. A typical RPA diagram is shown
in Fig.1(b). Although there have been calculations with
certain interactions which show quantitative agreement with empirical
values, there have been complete second-order perturbation calculations by
Siegel and Zamick \cite{sz} and Ellis and Siegel \cite{es}
which yielded very small polarization charges.
Evidently the non-RPA diagrams cancel with the RPA diagrams.
More recently, Skouras and M\"{u}ther \cite{sm} also
obtained small polarization charges in higher-order calculations
with G-matrices evaluated from Bonn potentials \cite{bonn}.

At the extreme it is known that for a closed $LS$ shell plus or minus
one particle, the first-order perturbation-theory corrections for the
magnetic dipole (M1) operator and to the Gamow-Teller (GT) operator
[$\sqrt{\frac{3}{4\pi}} (g_l\vec{l} + g_s\vec{s})$ and
$\sigma t_{\pm}$ respectively] vanish. One has to go to
higher-order perturbation theory and invoke exchange currents in order
to explain deviations from the lowest order.

In this work, we will study higher-shell corrections to properties
of systems involving one and two valence nucleons or holes.
Although the first-order perturbation-theory correction vanishes for
systems with one valence particle or hole,
we may get some effects for systems with two valence particles or
holes. For example, the effective interaction between the two valence nucleons
or holes can be renormalized (see, for example, Fig.2)
and hence the two-body wave function can change.

Rather than perform perturbative calculations, we will perform
matrix diagonalizations using the OXBASH shell-model (SM) code \cite{oxbash}.
The model space will consist of one or two valence particles or holes, plus
excitations of the core through ``$2\hbar\omega$''. By this we mean
that either two core-nucleons will be excited through one major shell
or one core-nucleon will be excited through two major shells.

There have been previous SM
calculations with ``$2\hbar\omega$'' corrections. For example,
van Hees {\it et al} \cite{vanH} and Wolters {\it et al} \cite{wolters}
have extended the classic ``$0\hbar\omega$''
calculations of Cohen and Kurath \cite{ck} by including
higher shells. At the time of this writing, another paper has
appeared \cite{jan}.
Whereas the above works have used a phenomenological
interaction and focused on getting a global fit to
experiment --- certainly a worthwhile quest --- we are more concerned
with seeing how well the G matrices, derived from modern realistic
nucleon-nucleon (NN) potentials,
are able to account for the measured properties.
Furthermore, we want to see how the various parts of the NN
interaction, especially the tensor and spin-orbit components, affect
the physical observables.

The calculations are performed with Brueckner G-matrices \cite{bruc}
calculated from a new Nijmegen potential (NijmII) \cite{nijm}.
Comparisons with other potentials, including the Hamada-Johnson
hard-core \cite{hj}, the original Reid-soft-core \cite{reid}
and a new Reid-soft-core (Reid93) \cite{nijm}
potential, will be made on a selected basis.
It should be emphasized that our calculations will not assume an
inert core. The single-particle energies are implicitly
generated in OXBASH \cite{oxbash} from the two-body
G-matrix elements as well as the one-body kinetic energy.
More explicitly, the matrix diagonalization will be performed
for the SM Hamiltonian
\begin{equation}
H_{\rm SM} = \left( \sum_{i=1}^A t_i -T_{\rm CM}\right)
   + \sum_{i<j}^A G_{ij} + \lambda (H_{\rm CM} - \frac{3}{2}\hbar\omega),
				\label{hsm}
\end{equation}
where $t$ and $T_{\rm CM}$ are the one-body and
center-of-mass (CM) kinetic energies respectively.
The G matrix is calculated according to
\begin{equation}
G(E_s) = v_{12} + v_{12} \frac{Q}{E_s - (h_1+h_2+v_{12})}v_{12}, \label{g}
\end{equation}
with $E_s$ the starting energy and
$h_i=t_i+u_i=\frac{\vec{p}^2_i}{2m} + \frac{1}{2}m\omega^2r_i^2$ the
one-body harmonic-oscillator Hamiltonian. We choose
$\hbar\omega$=16MeV for $A$=5 and 6 and $\hbar\omega$=14MeV for
$A$=14, 15, 17 and 18. The Pauli operator $Q$ in Eq.(\ref{g})
is defined to exclude the scattering of the
two particles in the ladder diagrams into the occupied states
as well as the states that will be included in the model space.
For example, in the ladder diagrams, the scattering into an
intermediate state with two nucleons in the $1s$-$0d$ major shells or
a state with one nucleon in the $0p$ shell and the other in the
$0f$-$1p$ shell is forbidden ($Q$=0), since these states
will be included in our $(0+2)\hbar\omega$ shell
model diagonalization. The starting energy $E_s$ in the G-matrix (\ref{g})
is taken to be
\begin{equation}
E_s = \epsilon_a+\epsilon_b+\Delta,
\end{equation}
where ($\epsilon_a+\epsilon_b$) is the unperturbed
energy for the initial two-particle state in the ladder diagrams
[$\epsilon_i = (2n_i+l_i+\frac{3}{2})\hbar\omega$ with $i=a,b$]
and $\Delta$, roughly speaking,
represents the interaction energy which in this work
is treated as an adjustable parameter whose value is
chosen to yield a reasonable nuclear binding energy.

Note that the G-matrix elements used in the $0\hbar\omega$
calculation will be the same as those used in the
$(0+2)\hbar\omega$ calculation so that the difference in the results
is solely due to the core-renormalization effects.

We force, for the lower-lying states,
the spurious CM motion
to be in its lowest $0\hbar\omega$ configuration by adding to the
SM Hamiltonian the last term in Eq.(\ref{hsm}) with $\lambda \gg 1$.

\section{Results for systems with one valence particle or hole}

We give results using the G-matrices
from the Nijmegen NN potential \cite{nijm} for $A$=5, 6, 14, 15, 17 and 18 in
Table I. The data in moments is taken from the compilations of
Ajzenberg-Selove \cite{ajz} and Raghavan \cite{data}.

We emphasize again that the single-particle energies
in our SM studies are not taken as parameters, rather, they are
implicitly generated by the kinetic energy and the G-matrix.
In a $(0+2)\hbar\omega$ space, the calculated ``single-particle'' splitting
between the $1/2^-_1$ and $3/2^-_1$ states in $^5\mbox{He}$ is
3.633 MeV. In $^{15}\mbox{O}$, the
$3/2^-_1$--$1/2^-_1$ splitting is 5.479 MeV.
The splitting between the $3/2^+_1$ and $5/2^+_1$ states in
$^{17}\mbox{O}$ is 6.260 MeV.

Focusing first on the one-valence-nucleon system $A$=17,
we see that indeed the values of $B({\rm GT})$ and $B({\rm M1})$
are nearly the same in the $0\hbar\omega$ space, in which
the configuration is a $d_{5/2}$ nucleon outside a closed
$^{16}\mbox{O}$ core, as in the larger $(0+2)\hbar\omega$ space.
The value of $B({\rm GT})$ [our definition is such that the factor
$(1.251)^2$ is {\em not} included; the operator is
$\sum_i \sigma(i)t_{\pm}(i)$] in the small $0\hbar\omega$ space
is 1.400 and in the large $(0+2)\hbar\omega$
space 1.381. For $A$=15, the story is the same. The value of $B({\rm GT})$
changes very little (from 0.333 to 0.326) in going from
the $0\hbar\omega$ to the $(0+2)\hbar\omega$ space.
Underlying this behavior for $A$=15 and $A$=17 is the previously mentioned
theorem that there are no first-order corrections to the GT or M1
moment for a closed $LS$ shell plus or minus one particle.
The small deviations are due to the fact that the shell
model goes beyond first order.

We next consider the E2 properties for $A$=17.
In the simple SM picture, the quadrupole
moment of $^{17}\mbox{O}$ is zero because the valence nucleon is a neutron.
In first-order perturbation theory, there is a contribution
due to the excitation of a core proton
through two major shells,
e.g. from $0s$ to $1s$-$0d$ or from $0p$ to $0f$-$1p$.
In the SM calculation that we do here, that effect
is implicitly taken into account, but also one has the excitation of
two core nucleons through one major shell, i.e., $(0p)^{-2} (1s0d)^2$.
These configurations contribute only
in second-order perturbation theory.

The calculated polarization charge for the $d_{5/2}$-neutron is the ratio
of the quadrupole moment of $^{17}\mbox{O}$ to that of a
$d_{5/2}$-proton which can be read off from Table I:
\begin{equation}
\epsilon_{\nu} = \frac{-0.720}{-5.924} = 0.123.
\end{equation}
Similarly, for the valence proton,
\begin{equation}
\epsilon_{\pi} = \frac{-6.183-(-5.924)}{-5.924} = 0.044.
\end{equation}
These are much smaller than the phenomenological values of 0.5 and
0.2.
These results agree qualitatively with
the higher-order perturbation theory results in Ref.\cite{sm}
where small polarization charges are also obtained.

To check if the second-order contributions are canceling the
first-order ones, we consider the first-order perturbation theory.
If we use an energy denominator of (--$2\hbar\omega$) for Fig.1(a),
the polarization charge for the neutron is
$\epsilon_{\nu} = 0.385$ and that for the proton is
$\epsilon_{\pi} = 0.131$.
However, when the calculated single-particle energy splittings
are used, the values are
$\epsilon_{\nu} = 0.263$ and $\epsilon_{\pi} = 0.087$.
These are still larger than the SM values of 0.123 and 0.044.

These discrepancies are partly due to the fact that in
the above first-order perturbation-theory approach,
the wave function is not normalized. The ground-state
wave function of $^6\mbox{Li}$ in the $(0+2)\hbar\omega$ SM calculation is
\begin{equation}
|{\rm g.s.}\rangle =
  68.446\% |1p\!-\!0h\rangle
+ 27.727\% |2p\!-\!1h\rangle + 3.827\% |3p\!-\!2h\rangle.
\end{equation}
A properly normalized wave function in the
perturbation theory would therefore lead to an approximately 32\%'s reduction
(from 0.263 to 0.18) in the first-order contribution to the effective
charge. Further reductions are caused by the higher-order effects
that are included in the $(0+2)\hbar\omega$ SM calculation.

We thus see that one must exhibit care when employing the
perturbation-theory approach to calculate effective operators.
The first-order perturbation theory first appears to be accountable
for the large empirical polarization charge
of nearly 0.5 for the $d_{5/2}$-neutron in $^{17}\mbox{O}$,
a more careful analysis involving the use of the self-consistent
single-particle energies
and the normalized wave function results in a factor of two's reduction
in $\epsilon_{\nu}$, leading to a much smaller result (0.385 to 0.18).
The $(0+2)\hbar\omega$ SM value of 0.123 is more consistent with the refined
first-order perturbation-theory result of 0.18, both
are smaller than the empirical value.
The large quadrupole moment observed for
$^{17}\mbox{O}$ is due to the fact the ``spherical core'' is actually
deformed and carries a non-zero angular momentum.
To give sufficiently large polarization charges, bigger than
$(0+2)\hbar\omega$ calculations are needed which will simulate
highly deformed admixtures into the ground state.

\section{Higher-shell effects for systems with
	 two valence nucleons (or holes)}

For $A$=14, we have a striking example of a case where there is a
large effect due to higher-shell admixtures, while
there is almost none for a one-valence-nucleon system.
The value of $B({\rm GT})$ for the decay
$^{14}\mbox{C} (J=0^+, T=1) \rightarrow ^{14}\mbox{N} (J=1^+, T=0)$
changes from 2.627 to 1.305. The value of $B(\rm M1)$
changes by almost the same ratio; indeed from isospin consideration the
spin part of $B({\rm M1})$ and $B({\rm GT})$ must change by exactly
the same ratio. The large change has been previously noted by us \cite{ann}.

What must be happening is that the wave function of the two valence nucleons
is changing due to a renormalization of the effective
interactions from higher shells. A typical second-order diagram
is shown in Fig.2.
We recall that this weak beta decay
is famous because, although it is {\it a prior} an allowed
GT transition, the life-time for the transition is very long,
indicating a near vanishing GT matrix element.

Furthermore, it was noted by Inglis \cite{inglis}
that in a two-hole calculation one
needs a tensor interaction to get a vanishing GT matrix element.
In our calculation we obtain a $B({\rm GT})$ value of 1.305.
We believe that the bare tensor interaction is too strong
so if we imagine turning on the tensor interaction from zero to its
full (bare) value, we get an overshoot. The matrix element
first gets
smaller in magnitude, then goes through zero, changes sign and gets
large in magnitude again. The effect of the
higher-shell admixtures is to effectively weaken the tensor
interaction in the valence space, thus causing the GT matrix element to
decrease in magnitude.

For $A$=6, there are some encouraging results from higher-shell effects.
The experimental magnetic dipole moment and electric quadrupole moment of the
$1^+$ ground state of $^6\mbox{Li}$ are respectively
$0.822 \mu_N$ and --$0.082 e\, {\rm fm^2}$ \cite{coon}. The sign of the
$^6\mbox{Li}$ quadrupole moment is opposite to that of the deuteron
$(Q=+0.28 e\, {\rm fm^2})$.
In the $LS$ limit,
the ground state of $^6\mbox{Li}$ has a $p^2$ configuration with
$L$=0, $S$=1, $J$=1. In that limit, the M1 moment is
$\mu = \mu_n+\mu_p = (-1.913 + 2.793)\mu_N = 0.880\mu_N$
and the quadrupole moment is zero.

In the small space, the results for $\mu$ and $Q$ are respectively 0.866$\mu_N$
and --0.237$e\, {\rm fm}^2$. In the large space, the results are
0.849$\mu_N$ and --0.163$e\, {\rm fm}^2$ in both cases closer to
experiment.

In Table II we give the results of the $jj$ and $LS$ limits for the
M1 and E2 moments of the $1^+$ ground state of
$^6\mbox{Li}$. For the M1 moment, these two limits
yield respectively $\mu=\frac{(\mu_p+\mu_n+1)}{3}=0.627\mu_N$ and
$\mu=\mu_p+\mu_n=0.880\mu_N$.
The experimental value 0.822$\mu_N$ lies between these two extremes.
The quadrupole moment $Q$ is zero in the $LS$ limit
($L$=0, $S$=1, $J$=1) and is positive in the $jj$ limit.
The expression for $Q$ in the $jj$ limit can be related to that of a
$p_{3/2}$ proton via the Wigner-Eckart theorem in terms of
Clebsh-Gordan and unitary Racah coefficients:
{\small
\begin{eqnarray}
Q((p^2_{3/2})^{J=1,T=0})
&=& (e_{\nu}+e_{\pi})(2\; 1\; 0\; 1|1\; 1)
\frac{U(2\; 3/2\; 1\; 3/2;\; 3/2\; 1)}{(2\; 3/2\; 0\; 3/2|3/2\; 3/2)}
Q(p_{3/2,\pi}) \nonumber \\
&=& -0.4 Q(p_{3/2,\pi}).
\end{eqnarray}
}
The quadrupole moment of a $j=l+1/2$ proton is $-l\hbar/(m\omega)$.
Hence the quadrupole moment for a $p_{3/2}$-proton is negative:
$Q(p_{3/2,\pi}) = -\hbar/(m\omega)$.
The $(p^2_{3/2})^{J=1,T=0}$ quadrupole moment is therefore
positive (1.04 $e\,{\rm fm}^2$ when $\hbar\omega=16$ MeV).

To get further insight into the $A$=6 behavior, we use a schematic
interaction
\begin{equation}
V_{\rm sche}=V_c+xV_{so}+yV_t.
\end{equation}
With $x$=$y$=1, this interaction gives
relative matrix elements similar to those of Bonn A \cite{bonn}
for the region of
$^{16}\mbox{O}$ \cite{muther}. Only partial waves with
a relative angular momentum $l$=0 and $l$=1 for the diagonal
and $l$=0 to $l$=2 for the off-diagonal are included \cite{ann}.
It is constructed so that it is easy to control
the strengths of the spin-orbit and tensor interactions by adjusting
$x$ and $y$, e.g., for $x$=0, we turn off the spin-orbit
interaction and for $y$=0, we turn off the tensor interaction.
The results for the schematic interaction are also presented in
Table II. The matrix elements of the interaction have been multiplied by
1.15 since they were fitted to $^{16}\mbox{O}$ but are being applied
to $A$=6.

We verify that for $x$=$y$=0, one gets the $LS$ wave function
with $L$=0, $S$=1 for the ground state of $^6\mbox{Li}$.
The magnetic moment in both the $0\hbar\omega$ and
$(0+2)\hbar\omega$ calculations
is $\mu_n+\mu_p$=0.880$\mu_N$ and the quadrupole moment is zero. These
results transcend the truncated shell model.

Note that the spin-orbit interaction and the tensor interaction
have opposite effects on the quadrupole moment. In the small space,
when the spin-orbit interaction is turned on at full strength
and the tensor interaction is turned off (i.e., $x$=1, $y$=0),
the quadrupole moment $Q$ is 0.121$e\, {\rm fm}^2$.
This is consistent with the fact that making the spin-orbit
interaction stronger takes one towards the $jj$ limit of Table II.

In the opposite extreme with
the tensor interaction on and the spin-orbit interaction off
(i.e., $x$=0, $y$=1), the value of $Q$ is --0.608$e\, {\rm fm}^2$.
For $x$=$y$=1, the value of $Q$ is --0.281$e\, {\rm fm}^2$.
In a previous work \cite{zzm}, we argued that the effective spin-orbit
interaction in the $p$ shell should be even stronger
than that given by the relativistic Bonn A with $m^*/m$=1 \cite{bonn}.
A reasonable set of parameters is $x$=1.5, $y$=1 with which
the small-space value of $Q$ is --0.114$e\, {\rm fm}^2$,
in reasonable agreement with experiment.
The main point here, however, is that the quadrupole moment
of $^6\mbox{Li}$
is extremely sensitive to the relative strengths of the spin-orbit
and tensor interactions in the nucleus.

As seen in Table I, the $2\hbar\omega$ admixtures enhance the quadrupole
properties of $A$=5, as expected. But but for $A$=6, the
magnitude of the quadrupole moment decreases from
--0.255 to --0.175$e\,{\rm fm}^2$ when
$2\hbar\omega$ configurations are included. We believe that for $A$=6
there are two
opposing tendencies. The building up of the effective charge
would cause the magnitude of $Q$ to increase. But the
``self weakening'' of the tensor interaction due to higher-order admixtures
will cause $Q$ to go from negative towards positive values.

A cluster approach for $^6\mbox{Li}$, i.e., an alpha-deuteron
cluster model calculation has been carried out by
Eskondarian {\it et al} \cite{esk}. Without
going into too much detail we focus on their results.
Whereas we get the magnetic moment of $^6\mbox{Li}$
larger than experiment, they get it smaller.
Whereas we get the quadrupole moment of $^6\mbox{Li}$
to be negative, they get it positive. It is not
clear if the differences are due to the differences between
the SM and the cluster model, or in the differences between the
interactions that are used. The SM will also yield
a positive quadrupole moment if
the spin-orbit interaction is stronger and/or the tensor interaction is
weaker than what we have. Another possibility to be explored is that
the cluster model has correlations which require even larger SM
calculations than the ``$2\hbar\omega$'' ones performed here.

\section{Closing remarks}
We have shown not only that higher-shell effects are important for
nuclear properties but that they enter in a nonuniform way as we add
nucleons to the nucleus. For example, higher-shell effects are much
larger for the spin properties of a two-valence-particle (or hole)
system than they are for a one-valence-particle (or hole) system.
As we go from a small space ($0\hbar\omega$) to a large space
[$(0+2)\hbar\omega$], the
value of B(GT) for $A$=15 experiences little change (from 0.333 to 0.326),
but for $A$=14 the change is much bigger (from 2.627 to 1.305).
This can be understood by
the fact that the effective interaction between the two valence holes gets
renormalized. In this renormalization, the effective tensor interaction
(acting in the valence space) becomes weaker.

The $A$=6 and $A$=14 cases offer good tests of the effective interaction.
For $^6\mbox{Li}$, which has often been proposed as an
ideal isoscalar target (see Ref.\cite{coon} for more details),
the results for the J=$1^{+}$ static
quadrupole moment and magnetic moment are very sensitive to the
details of the spin-orbit and tensor interactions that are used. There
is a surprisingly large scatter of theoretical results --- both positive
and negative values for the quadrupole moment of the J=$1^{+}$ ground
state of $^6\mbox{Li}$ have been obtained. We note that the tensor
interaction gives a negative value and the spin-orbit interaction a
positive value for this quantity. Our results with the effective interactions
calculated from the Nijmegen potential NijmII
as applied to ``two-particle'' systems are encouraging in
the sense that higher-shell effects consistently go in the right
direction. There is still a problem, however, in that the one-body
quadrupole polarization charges come out too small.

If we want to use the nucleus as a laboratory for fundamental studies,
e.g., of fundamental symmetries, applications to astrophysics, or, as we
have emphasized here, the effective interaction between two nucleons in
a nucleus, we must include higher-shell effects. Indeed, our work here
represents only a step in the right direction. In the future,
excitations beyond '2$\hbar\omega$' will have to be included.
Still, it is encouraging that we are able to get a better
systematic feeling for how higher-shell admixtures affect
nuclear properties in a nuclear medium.

\section*{Acknowledgment}
This work was supported by the National Science Foundation,
Grant No. PHY-9103011 (D.C.Z.) and
the U.S. Department of Energy under Grant No. DE-FG05-86ER-40299,
Division of High Energy and Nuclear Physics (L.Z. and M.F.).

\pagebreak

\begin{small}

\end{small}

\pagebreak

\begin{small}

\noindent
{\bf Table I}.
The static and transitional properties in selected nuclei with
one or two valence nucleons (or holes) in
the $0\hbar\omega$ and $(0+2)\hbar\omega$ shell-model calculations
using G-matrices calculated from a new Nijmegen
local NN potential \cite{nijm}. $B({\rm M1})$ is in units of $\mu_N^2$,
$\mu$ in units of $\mu_N$ and $Q$ in units of $e\, {\rm fm}^2$.
Experimental data are taken from Refs.\cite{ajz,data}.
\begin{center}
\begin{tabular}{l|l|cc|c}\hline\hline
Nucleus & Quantity & $0\hbar\omega$ & $(0+2)\hbar\omega$ & Expt \\ \hline
$A$=5 & $B({\rm GT}) (3/2^- \rightarrow 3/2^-)$&  1.666 &  1.632 & \\
      & $\mu (^5\mbox{He})$                    & -1.913 & -1.864 & \\
      & $\mu (^5\mbox{Li})$                    &  3.793 &  3.741 & \\
      & $Q(^5\mbox{He})$                       & -2.592 & -2.722 & \\
      & $Q(^5\mbox{Li})$                       &  0.000 & -0.303 & \\
								\hline
$A$=6 & $B({\rm GT}) (0^+1 \rightarrow 1^+0)$ &  5.626 &  5.397 & 4.847 \\
      & $B({\rm M1}) (0^+1 \rightarrow 1^+0)$ &  16.29 &  15.57 & \\
      & $\mu (1^+0)$                          &  0.866 &  0.849 & 0.822\\
      & $Q (1^+0)$                            & -0.255 & -0.175 & -0.082\\
								\hline
$A$=14& $B({\rm GT}) (0^+1 \rightarrow 1^+0)$ &  2.627 &  1.305 & 0.000\\
      & $B({\rm M1}) (0^+1 \rightarrow 1^+0)$ &  6.916 &  3.842 & \\
      & $\mu (1^+0)$                          &  0.634 &  0.499 & 0.404\\
      & $Q (1^+0)$                            &  1.761 &  2.164 & 1.56\\ \hline
$A$=15& $B({\rm GT}) (1/2^-\rightarrow 1/2^-)$& 0.333  &  0.326 & 0.270 \\
      & $\mu (^{15}\mbox{N})$                 &-0.264  & -0.276 & -0.283\\
      & $\mu (^{15}\mbox{O})$                 & 0.638  &  0.654 & 0.719\\
								\hline
$A$=17& $B({\rm GT})(5/2^+ \rightarrow 5/2^+)$&  1.400 &  1.381 & 1.089 \\
      & $\mu (^{17}\mbox{O})$                 & -1.913 & -1.869 & -1.894 \\
      & $\mu (^{17}\mbox{F})$                 &  4.793 &  4.748 & 4.722\\
      & $Q (^{17}\mbox{O})$                   &  0.000 & -0.720 & -2.578 \\
      & $Q (^{17}\mbox{F})$                   & -5.924 & -6.183 & -5.8\\ \hline
$A$=18& $B({\rm GT}) (0^+1 \rightarrow 1^+0)$ &  5.312 &  5.436 & 3.301 \\
      & $B({\rm M1}) (0^+1 \rightarrow 1^+0)$ &  16.75 &  16.19 & \\
      & $\mu (1^+0)$                          &  0.858 &   ---  & \\
      & $Q (1^+0)$                            & -0.514 &   ---  & \\
							\hline\hline
\end{tabular}
\end{center}
\end{small}

\pagebreak

\begin{small}

\noindent
{\bf Table II}.
The magnetic dipole moment $\mu$
(in units of $\mu_N$) and quadrupole moment $Q$ (in units of $e\, {\rm fm}^2$)
for the ground state of $^6\mbox{Li}$ in
the $0\hbar\omega$ and $(0+2)\hbar\omega$ shell-model calculations
using (1) the schematic interaction with varying spin-orbit
($x$) and tensor ($y$) strengths, and (2) G-matrices calculated from
different realistic NN potentials: Hamada-Johnson \cite{hj},
Reid-soft-core \cite{reid}, new Reid-soft-core (Reid93) \cite{nijm},
and new Nijmegen \cite{nijm}.

\begin{center}
\begin{tabular}{cc|cc|cc}\hline\hline
\multicolumn{2}{c|}{Interaction} & \multicolumn{2}{c|}{$\mu$}
			& \multicolumn{2}{c}{$Q$} \\ \hline
$x$ & $y$ & $0\hbar\omega$ & $(0+2)\hbar\omega$
          & $0\hbar\omega$ & $(0+2)\hbar\omega$ \\ \hline
  0.0 & 0.0 & 0.880 & 0.880 & 0.000 &  0.000 \\
  0.0 & 0.5 & 0.867 & 0.859 &-0.345 & -0.415 \\
  0.0 & 1.0 & 0.840 & 0.816 &-0.608 & -0.711 \\ \hline
  0.5 & 0.0 & 0.877 & 0.878 &+0.038 & +0.033 \\
  0.5 & 0.5 & 0.872 & 0.866 &-0.221 & -0.276 \\
  0.5 & 1.0 & 0.856 & 0.835 &-0.446 & -0.533 \\ \hline
  1.0 & 0.0 & 0.869 & 0.871 &+0.121 & +0.106 \\
  1.0 & 0.5 & 0.870 & 0.866 &-0.085 & -0.135 \\
  1.0 & 1.0 & 0.863 & 0.845 &-0.281 & -0.358 \\ \hline
  1.5 & 0.0 & 0.854 & 0.859 &+0.221 & +0.195 \\
  1.5 & 0.5 & 0.861 & 0.860 &+0.055 & +0.006 \\
  1.5 & 1.0 & 0.860 & 0.848 &-0.114 & -0.185 \\ \hline
\multicolumn{2}{c|}{$LS$ limit}     & \multicolumn{2}{c|}{0.880}
				    & \multicolumn{2}{c}{0.000} \\
\multicolumn{2}{c|}{$jj$ limit}     & \multicolumn{2}{c|}{0.627}
				    & \multicolumn{2}{c}{1.037} \\ \hline
\multicolumn{2}{c|}{Hamada-Johnson} & 0.861 & 0.840 & -0.364 & -0.285 \\
\multicolumn{2}{c|}{Reid-soft-core} & 0.862 & 0.842 & -0.307 & -0.224 \\
\multicolumn{2}{c|}{Reid93}         & 0.866 & 0.848 & -0.255 & -0.181 \\
\multicolumn{2}{c|}{NijmII}         & 0.866 & 0.849 & -0.255 & -0.175 \\
					\hline\hline
\end{tabular}
\end{center}
\end{small}

\vspace{0.5in}

\section*{Figure caption}

\vspace{0.2in}

\noindent
{\bf Fig.1} Core-polarization diagrams for E2 effective charge.
	(a) First-order diagram; (b) Typical RPA graph.

\vspace{0.2in}

\noindent
{\bf Fig.2} Core renormalization of the effective interaction.
This changes the two-particle wavefunction and hence static properties
like M1 and E2 moments.


\begin{thebibliography}{99}
\bibitem{sz} S. Siegel and L. Zamick, Phys. Lett. {\bf B28}, 453(1969);
    	     S. Siegel and L. Zamick, Nucl. Phys. {\bf A145}, 89(1970).
\bibitem{es} P.J. Ellis and S. Siegel, Phys. Lett. {\bf B34}, 177(1971).
\bibitem{sm} L.D. Skouras and H. M\"{u}ther, Nucl. Phys. {\bf A534}, 128(1991).
\bibitem{bonn} R. Machleidt, Adv. in Nucl. Phys. Vol. 19, 189(1989).
\bibitem{oxbash} A. Etchegoyen, W.D.M. Rae, N.S. Godwin, B.A. Brown,
	W.E. Ormand, and J.S. Winfield, the Oxford--Buenos Aires --- MSU
	Shell Model Code (OXBASH) (unpublished).
\bibitem{vanH} A.G.M. van Hees and P.W.M. Glaudemans,
	Z. Phys. {\bf A314}, 323(1983); {\it ibid} {\bf A315}, 223(1984).
\bibitem{wolters} A.A. Wolters, A.G.M. van Hees and P.W.M. Glaudemans,
	Europhysics Lett. {\bf 5(1)}, 7(1988).
\bibitem{ck} S. Cohen and D. Kurath, Nucl. Phys. {\bf 73}, 1(1965).
\bibitem{jan}H. Nakada and T. Otsuka, ``E2 properties of nuclei far from
	stability and the proton-halo problem of $^8\mbox{B}$,''
	preprint, (LANL Electronic Preprint Library, nucl-th/9311004).
\bibitem{bruc} K.A. Brueckner, Phys. Rev. {\bf 97}, 1353(1955);
		 {\bf 100}, 36(1955).
\bibitem{nijm} J.J. de Swart and M. Rentmeester, private communication.
\bibitem{hj} T. Hamada and I.D. Johnson, Nucl. Phys. {\bf 34}, 382(1962).
\bibitem{reid} R.V. Reid, Ann. of Phys. {\bf 50}, 411(1968).
\bibitem{ajz} F. Ajzenberg-Selove,
	Nucl. Phys. {\bf A449}, 1(1986);
	Nucl. Phys. {\bf A460}, 1(1986);
	Nucl. Phys. {\bf A475}, 1(1986);
	Nucl. Phys. {\bf A490}, 1(1988).
\bibitem{data} P. Raghavan, Atomic Data and Nuclear Data Tables,
	{\bf 42}(2), 189(1989).
\bibitem{ann} D.C. Zheng and L. Zamick, Ann. Phys. (NY) {\bf 206}, 106(1991).
\bibitem{inglis} D.R. Inglis, Rev. Mod. Phys. {\bf 25}, 390(1953).
\bibitem{coon} N.W. Schellingerhout, L.P. Kok, S.A. Coon, and R.M. Adam,
	``Nucleon polarization in three-body models of polarized
	$^6\mbox{Li}$,'' preprint, to appear in the December 1993
	issue of Phys. Rev. C.
\bibitem{muther} H. M\"{u}ther, private communication.
\bibitem{zzm} L. Zamick, D.C. Zheng, and H. M\"{u}ther,
	Phys. Rev. {\bf C45}, 2763(1992);
\bibitem{esk} A. Eskandarian, D.R. Lehman and W.C. Parke,
	Phys. Rev. {\bf C38}, 2341(1988).
\end{thebibliography}
\end{document}